\documentclass[11pt]{article}
\usepackage[a4paper,margin=2.5cm]{geometry}
\usepackage{amsmath,amssymb,amsfonts}
\usepackage{authblk}
\usepackage{booktabs,longtable,array,multirow,makecell}
\usepackage{graphics}
\usepackage{float}
\usepackage{hyperref}
\usepackage{caption}
\usepackage{adjustbox}
\usepackage{setspace}
\usepackage{enumitem}
\usepackage{titlesec}
\titleformat{\section}{\large\bfseries}{\thesection.}{0.5em}{}
\titleformat{\subsection}{\normalsize\bfseries}{\thesubsection.}{0.5em}{}

\title{Comparison of Effective Dissipation Channels in Warm Higgs Inflation from Warm Background Evolution}

\author[1]{Wei Cheng}
\author[1]{Xin Peng} 
\author[2]{Jia-wei Zhang\thanks{ Corresponding authors. Email: chengwei@cqupt.edu.cn, pengxin@stu.cqupt.edu.cn, jwzhang@cqust.edu.cn, zhoury@cqupt.edu.cn, panyu@cqupt.edu.cn}}
\author[1]{Ruiyu Zhou$^{*}$}
\author[1]{Yu Pan$^{*}$}

\affil[1]{School of Electronic Science and Engineering, Chongqing University of Posts and Communications, Chongqing 400065, China }
\affil[2]{Department of Physics, Chongqing University of Science and Technology, Chongqing 401331, China} 

\date{} 

\begin{document}

\maketitle

\begin{abstract}
Within the framework of warm Higgs inflation, a systematic comparison is carried out among seven effective dissipation channels (EDC) constructed from combinations of the three basic dissipation channels, namely the low temperature (LT), high temperature (HT), and threshold (Th) channels. Adopting a unified treatment of warm background evolution, complexity penalization, and boundary consistency checks, the comparison is performed in terms of their distributions of the best fit points in ($n_s$, $r$) plane, relative BIC hierarchy, channel dominance patterns, and warmness indicators. The results show that, except for the pure HT EDC $\Upsilon_{\mathrm{010}}$, the best fit points of the other six EDC are clustered within a small region of the ($n_s$, $r$) plane, around $n_s \approx 0.965$ and $r \approx (3.68 \to 3.74)\times10^{-3}$. In contrast, $\Upsilon_{\mathrm{010}}$ is displaced from this main cluster, with a representative best fit point near $n_s = 0.9552$ and $r = 6.0\times10^{-3}$. Under both the unified scan and the 1200-point refined rescoring, the pure LT EDC $\Upsilon_{\mathrm{100}}$ remains top-ranked, while $\Upsilon_{\mathrm{011}}$ and $\Upsilon_{\mathrm{111}}$ remain disfavored, indicating that the overall hierarchy is stable under the present boundary check criterion. Warmness diagnostics further show that $\Upsilon_{\mathrm{100}}$ corresponds to $Q_* \approx 35.7$ and $T_*/H_* \approx 1.90\times10^{3}$, placing it in the strong warm regime, whereas $\Upsilon_{\mathrm{011}}$ gives $T_*/H_* \approx 0.31$, already below the warmness threshold. The channel fractions, boundary checks, and constrained internal-mixing probes consistently indicate that the best fit points of the multi-channel EDC do not form a stable internally mixed region, but instead lie closer to a single channel dominated regime. The hierarchy among EDC is therefore determined not by their positions in ($n_s$, $r$) plane, but rather by the complexity cost required to reach those positions, the dominance pattern, and the allowed warmness range.
\end{abstract}

\section{Introduction}

Inflation provides a unified explanation of the flatness problem, the horizon problem, and the origin of primordial perturbations in the early Universe, while high precision CMB measurements have compressed viable models into a rather narrow working window. The standard inflationary framework and the theory of primordial perturbations were established in the seminal works of refs \cite{Starobinsky:1980te, Guth:1980zm, Linde:1981mu, Mukhanov:1981xt}, and the Planck 2018 and BICEP/Keck data further restrict the scalar spectral index to a neighborhood of $n_s\simeq0.965$ while tightening the allowed range of the tensor to scalar ratio \cite{Planck:2018jri, BICEP:2021xfz}. Any viable model must therefore do more than merely reproduce an acceptable $n_s$ and a small $r$; it must also specify the dynamical mechanism by which these observables are realized.

For Higgs inflation with non-minimal coupling, the plateau potential in Einstein frame has been extensively studied in the cold inflation setting \cite{Bezrukov:2007ep, Rubio:2018ogq, Barrie:2021mwi, Cheng:2018axr,Haque:2025uis, Rubio:2026acm, Hertzberg:2010dc, Kaganovich:2025iht, Ema:2017loe,Salvio:2015kka}. Once warm inflation is introduced, however, the inflaton potential is no longer the only essential input. The dissipative coefficient and the thermal bath evolution also reshape the formation of observable quantities. In warm Higgs inflation, the potential shape, the non-minimal coupling, and the manner in which dissipation operates are entangled, so that the question of how to compare different effective channels for a given potential becomes a separate and necessary problem \cite{Cheng:2024uvn}.

Unlike cold inflation, warm inflation allows persistent dissipation and radiation production during the inflationary stage. The dissipative coefficient is therefore not a marginal correction, but a core ingredient entering simultaneously the background evolution, the maintenance of the thermal bath, and the generation of scalar perturbations. The basic framework of warm inflation was formulated by Berera and collaborators, and later reviews gave systematic accounts of its microscopic basis, finite-temperature field theory setting, and fluctuation dissipation relation \cite{Berera:1995wh, Berera:1995ie, Berera:1996nv, Berera:1996fm, Berera:2008ar,Yuennan:2024nje,Yuennan:2025szw}.

Within this framework, two directions have been actively pursued. One line of work focuses on how the scalar power spectrum, thermal occupation number, and radiation inflaton coupling enter $n_s$ and $r$, another attempts to derive dissipative coefficients in the low temperature (LT) or high temperature (HT) regime from microscopic interactions \cite{Taylor:2000ze, Hall:2003zp, Moss:2007cv, Moss:2008yb, Graham:2009bf, Sayar:2017pam, Setare:2013kja, Bastero-Gil:2004oun, Kamali:2023lzq, Das:2018rpg,Arya:2019wck, Das:2020xmh}. Together these studies make clear that, in warm inflation, the dissipative coefficient is not a posterior tuning term but a key quantity running through the full chain from background evolution and thermal bath to perturbation transfer.

Existing studies still tend to focus on constraining the parameters of a given dissipative form. Much less attention has been paid to whether a set of effective dissipation channels (EDC) can be ordered systematically under a unified comparison scheme. This issue becomes nontrivial when LT power law channels, HT power law channels, and threshold (Th) suppressed channels coexist. These EDC differ physically, yet may appear close at the phenomenological level, so that the comparison of EDC becomes a problem in its own right \cite{Bastero-Gil:2011rva, Bastero-Gil:2010dgy, Moss:2011qc, Ramos:2013nsa, Bastero-Gil:2012akf}.

Recent work has further shown that dissipative coefficients in the LT and HT regimes do not follow the same approximation logic. The Warm Little Inflaton scenario and its extensions provide protected realizations that are closer to the HT representative case. At the same time, if a broad scan is allowed in a unified power-law interface, the HT parameter region may algebraically cover LT-like corners. The distinction between a parameterization label and a genuine physical origin must therefore be kept explicit \cite{Bastero-Gil:2006ahd, Bastero-Gil:2018yen, Bastero-Gil:2016qru, Bastero-Gil:2018uep, Laine:2021ego}.

On the observational side, ACT analyses and forecasts for Simons Observatory and CMB-S4 continue to improve the discriminating power of early Universe data \cite{ACT:2020gnv, SimonsObservatory:2018koc, CMB-S4:2016ple}. Parameter constraints on warm inflation, concrete implementations of warm Higgs inflation, and Stage-IV forecasts have also advanced steadily, but most of these studies still proceed in the form of single model parameter analyses \cite{Cheng:2024uvn, Reyimuaji:2020bkm, Montefalcone:2022jfw, Rasouli:2018kvy, Panotopoulos:2015qwa, Benetti:2016jhf, Freese:2024ogj, Santos:2024plu}. Previous results already showed that the dissipative strength in warm Higgs inflation can substantially modify both the background trajectory and the observable predictions \cite{Cheng:2024uvn,Cheng:2025rmf}. Against this background, moving from ``parameter effects within a single dissipative form'' to ``a systematic ordering of EDC'' is a natural next step.

For this reason, the present work does not enter directly into the full space of continuous mixing weights. Instead, it first compares the seven EDC obtained by turning on or off three basic effective components in the warm Higgs plateau potential. This choice is not made merely for numerical control. More importantly, the set of EDC corresponds to an independent physical question: once the potential, the scoring prescription, and the search boundaries are all unified, do different operating modes form a stable and interpretable hierarchy? Even if several channels yield nearby best fit points in $(n_s,r)$ plane, they may still differ markedly in their dominant channels, warmness ranges, and complexity costs. The comparison of EDC is therefore worth addressing in its own right.

From the perspective of the existing literature, most studies remain centered on continuous parameter constraints for a single dissipative form. Whether several EDC admit a stable hierarchy under a unified criterion has received much less attention. What is done here is not yet another refit of one model, but a comparison among EDC themselves.

Accordingly, the focus is not on further improving the local fit quality of one model, but on asking whether the dominance pattern, complexity cost, and boundary stability of effective channels can form an interpretable hierarchy under a common implementation. The unified scan, complexity penalty, rescoring procedure, and boundary checks are therefore combined into a single analysis chain. The object of interest is not an isolated best fit point, but the differences among channels in their dominance pattern and robustness. The main novelty of this work lies in this change of comparison target and analysis chain, rather than in any claim of a final discrimination among microscopic dissipative mechanisms. The results should thus be understood as a comparison of EDC based on warm background outputs and proxy effective observables.

The remainder of this paper is organized as follows. Sec.~\ref{Sec2} presents the theoretical framework of the warm Higgs plateau potential, the basic dynamical equations, and the parametrization of the EDC. Sec.~\ref{Sec3} describes the numerical solution of the warm background trajectory, the determination of the end point of inflation and the pivot scale quantities, and the scoring criteria and boundary-check settings adopted in the subsequent comparison. Sec.~\ref{Sec4} reports the main results, including the distribution of best fit points in ($n_s$, $r$) plane, the relative BIC hierarchy, the 1200-point refined rescoring results, the dominant channel patterns, the warmness diagnostics, and the boundary checks together with the constrained internal-mixing probes. Sec.~\ref{Sec5} summarizes the main conclusions.

\section{Warm Higgs inflationary and EDC parameterization}\label{Sec2}

\subsection{Warm Higgs inflationary potential and background dynamics}

The warm inflation dynamics is discussed in the Einstein frame, and the canonically normalized effective inflaton field is denoted by $h$. For warm Higgs models with non-minimal coupling, let $\phi$ be the original Higgs mode in the Jordan frame. The Einstein frame potential can then be written as
\begin{equation}
U_E(\phi)=\frac{\lambda\,\phi^4}{4\left(1+\xi\phi^2/M_{\rm Pl}^2\right)^2}.
\end{equation}
In the large field slow roll regime, it is convenient to rewrite the potential in terms of the canonical field $h$ in the familiar plateau form
\begin{equation}
U(h)=\frac{\lambda M_{\rm Pl}^4}{4\xi^2}\left(1-e^{-\sqrt{2/3}\,h/M_{\rm Pl}}\right)^2,
\end{equation}
with the field redefinition
\begin{equation}
\left(\frac{dh}{d\phi}\right)^2=
\frac{1+\xi(1+6\xi)\phi^2/M_{\rm Pl}^2}
{\left(1+\xi\phi^2/M_{\rm Pl}^2\right)^2}.
\end{equation}
This form is standard for the warm Higgs plateau potential and makes it convenient to discuss the potential and EDC separately \cite{ACT:2020gnv, SimonsObservatory:2018koc, CMB-S4:2016ple}.

The essential difference between warm and cold inflation is that the inflaton and the radiation bath exchange energy continuously during inflation. The background equations must therefore be written as
\begin{equation}
3M_{\rm Pl}^2H^2=\rho_h+\rho_R, \qquad
\ddot h+(3H+\Upsilon)\dot h+U_{,h}=0, \qquad
\dot\rho_R+4H\rho_R=\Upsilon\dot h^2.
\end{equation}
Here $H$ is the Hubble rate, $\rho_h=\dot h^2/2+U(h)$ is the inflaton energy density, $\rho_R$ is the radiation energy density, and $\Upsilon$ is the dissipative coefficient. Dissipation strengthens the effective friction of the inflaton while continuously sourcing the radiation bath, so that the background trajectory, the thermal bath strength, and the subsequent perturbation spectrum are naturally coupled \cite{Berera:1995wh, Berera:1995ie, Berera:1998gx, Berera:1998px}.

\subsection{Slow roll approximation and warmness diagnostics}

In the slow roll regime of warm inflation, the inflaton kinetic energy and the radiation energy density are both subdominant to the potential, and their evolution is comparatively mild. The background equations then reduce to
\begin{equation}
3M_{\rm Pl}^2H^2\simeq U(h), \qquad
3H(1+Q)\dot h\simeq -U_{,h}, \qquad
\rho_R\simeq \frac{3}{4}Q\dot h^2 ,
\end{equation}
where
\begin{equation}
Q\equiv \frac{\Upsilon}{3H}
\end{equation}
is the dimensionless dissipative strength. When $Q\ll1$, the system lies in the weak dissipative regime. When $Q\gtrsim1$, it enters the strong dissipative regime. The quantity $Q_*$ at the pivot scale will be used below as one of the key diagnostics in comparing EDC \cite{Berera:2008ar, Hall:2003zp, Moss:2007cv, Bastero-Gil:2011rva}.

The potential slow roll parameters are defined by
\begin{equation}
\epsilon_V\equiv \frac{M_{\rm Pl}^2}{2}\left(\frac{U_{,h}}{U}\right)^2,\qquad
\eta_V\equiv M_{\rm Pl}^2\frac{U_{,hh}}{U},\qquad
\beta_\Upsilon\equiv M_{\rm Pl}^2\frac{U_{,h}}{U}\frac{\Upsilon_{,h}}{\Upsilon}.
\end{equation}
The warm slow roll conditions can be written as
$\epsilon_V,|\eta_V|,|\beta_\Upsilon|\ll1+Q$.
Besides $Q$, another central warmness variable is
\begin{equation}
\frac{T}{H}.
\end{equation}
At horizon crossing, the condition $T_*/H_*>1$ is required for a genuinely warm interpretation. In what follows, $\log_{10}(T_*/H_*)$ is treated as an explicit warmness diagnostic \cite{Berera:2008ar, Hall:2003zp, Moss:2008yb, Graham:2009bf}.

\subsection{Effective observables}
In numerical implementation, the full warm background trajectory is first obtained from the background equations, and both the end point of inflation and the pivot-scale quantities are determined from that background solution. The scalar spectral index and the tensor to scalar ratio entering the subsequent scoring step are then constructed from the potential slow roll quantities and the dissipative strength at the pivot scale. Specifically,
\begin{equation}
n_s \simeq 1-\frac{6\epsilon_{V,*}-2\eta_{V,*}}{1+Q_*},
\qquad
r \simeq \frac{16\epsilon_{V,*}}{(1+Q_*)^2}.
\end{equation}
At the same time, $Q_*$, $T_*/H_*$, and the channel fractions at the pivot scale are retained as diagnostics of warmness and channel dominance. The quantities passed to the subsequent scoring step are therefore effective observables constructed from the warm background outputs, the potential slow roll quantities, and the dissipative strength. The detailed determination of the background trajectory, the end of inflation, and the pivot scale quantities is described in Sec.~\ref{Process}.

\subsection{EDC}

The total dissipative coefficient is written as the sum of three components,
\begin{equation}
\Upsilon_{\mathrm{tot}}(h,T)=\Upsilon_{\rm LT}(h,T)+\Upsilon_{\rm HT}(h,T)+\Upsilon_{\rm Th}(h,T).
\end{equation}
The LT channel is taken as
\begin{equation}
\Upsilon_{\rm LT}(h,T)=C_{\rm LT}\frac{T^3}{\bar h^2},\qquad
\bar h\equiv\sqrt{h^2+h_0^2},
\end{equation}
the HT inspired channel as
\begin{equation}
\Upsilon_{\rm HT}(h,T)=C_{\rm HT}\,T\left(\frac{\bar h}{\mu}\right)^{\alpha_h}
\left(\frac{T}{\mu}\right)^{\alpha_T},
\end{equation}
and the Th channel as
\begin{equation}
\Upsilon_{\rm Th}(h,T)=C_{\rm Th}\,\mu
\left(\frac{|h|}{\mu}\right)^a
\left(\frac{T}{\mu}\right)^c
\exp\!\left[-\frac{M_{\rm eff}(h,T)}{T}\right],
\end{equation}
with
\begin{equation}
M_{\rm eff}^2(h,T)=(g|h|)^2+m_0^2+\kappa_TT^2.
\end{equation}

Seven EDC can be written as:
\begin{equation}
\Upsilon_{\mathrm{ijk}}(h,T)=\mathrm{i}\times\Upsilon_{\rm LT}(h,T)+\mathrm{j}\times\Upsilon_{\rm HT}(h,T)+\mathrm{k}\times\Upsilon_{\rm Th}(h,T).
\end{equation}
where $\mathrm{i,j,k} \in \{0,1\}$.

The purpose of this parameterization is to cover, within one common parameter dictionary, a LT power-law channel, a HT interface around a representative HT behaviour, and a Th suppressed EDC, so that different EDC can be compared within a single numerical framework \cite{Moss:2007cv, Bastero-Gil:2011rva, Bastero-Gil:2010dgy, Ramos:2013nsa, Laine:2021ego}.

In the present work the HT channel is written as $\Upsilon_{\rm HT}\propto \bar h^{\alpha_h}T^{1+\alpha_T}$. The physical idea is to keep the leading linear dependence on $T$ as the dominant HT behaviour and to allow small deviations around it through $\alpha_h$ and $\alpha_T$. The HT channel should therefore be regarded not as an arbitrary power law, but as a HT inspired form whose leading behaviour is controlled by $T$.

In the actual scan, the allowed ranges of $\alpha_h$ and $\alpha_T$ are restricted to finite intervals. This retains enough freedom to capture the shape of the HT channel while avoiding an overly broad parameterization. The HT channel in this paper is therefore best understood as an effective interface built around a representative high-temperature behaviour, rather than as direct evidence in favour of a unique microscopic dissipative origin. If a best fit point lands in a region that resembles the LT form, it is interpreted here simply as an overlap within the present effective parameterization.

These effective channel parameterizations provide the physical input required for comparing EDC, but by themselves they do not determine the relative hierarchy. To turn this comparison into a computable problem, the next section specifies a unified set of search boundaries, scoring prescriptions, and diagnostics.

\section{Warm background integration, numerical implementation, and scope of applicability}\label{Sec3}

To place the comparison of EDC on a common basis of physical input, statistical definition, and robustness tests, this section first describes the numerical solution of the warm background trajectory together with the determination of the end of inflation and of the pivot scale quantities, and then explains the scoring prescription, the definition of the data term, and the scope of validity of the results. All rankings and robustness checks discussed later are based on this unified setup.

\begin{table}[H]
\centering
\caption{Search boundaries of the continuous parameters explicitly implemented in the present numerical setup.}
\begin{adjustbox}{max width=\textwidth}
\begin{tabular}{lll}
\toprule
Parameter & Physical meaning & unified scan range\\
\midrule
$\lambda$ & Einstein frame plateau potential coupling & $[10^{-7},\,10^{-1}]$ \\
$\xi$ & non-minimal coupling parameter & $[1,\,10^{7}]$ \\
$\log_{10} C_{\rm LT}$ & LT-channel amplitude & $[-12,\,8]$ \\
$\log_{10} C_{\rm HT}$ & HT-channel amplitude & $[-12,\,8]$ \\
$\log_{10} C_{\rm Th}$ & threshold-channel amplitude & $[-16,\,8]$ \\
$g$ & threshold effective mass coupling & $[10^{-5},\,1.5]$ \\
$a$ & field-power index in the threshold channel & $[-4,\,6]$ \\
$c$ & temperature power index in the threshold channel & $[-2,\,6]$ \\
$\alpha_h,\alpha_T$ & HT power law indices & $[-1.2,\,1.2]$ \\
$\log_{10} h_0$ & regularization scale in $\bar h=\sqrt{h^2+h_0^2}$ & $[-4,\,1]$, i.e. $h_0\in[10^{-4},10]$ \\
$\log_{10}\mu$ & reference scale in the HT/threshold channels & $[-4,\,1]$, i.e. $\mu\in[10^{-4},10]$ \\
$\log_{10} m_0$ & zero-temperature mass term in the threshold channel & $[-3,\,2]$, i.e. $m_0\in[10^{-3},10^{2}]$ \\
$\kappa_T$ & thermal-mass correction coefficient & $[0,\,30]$ \\
\bottomrule
\end{tabular}\label{Range}
\end{adjustbox}
\end{table}

\noindent Tab.~\ref{Range} lists the search boundaries explicitly hard coded in the present numerical implementation. The ``unified scan range'' refers to the first full parameter scan over the seven EDC under this common set of search boundaries, a common scoring prescription, and a common diagnostic setup. This first scan yields the best candidate point for each channel and serves as the starting point for the hierarchy comparison discussed below.

\subsection{Warm background integration, the end of inflation, and the pivot scale}\label{Process}

For a given set of potential parameters and channel parameters, the coupled inflaton radiation system is first integrated numerically on a homogeneous FRW background. The dynamical variables are chosen as $y(t)=\bigl(h,\dot h,\rho_R,N\bigr)$, where $h$ is the canonically normalized inflaton in the Einstein frame, $\rho_R$ is the radiation energy density, and $N$ is the number of e-folds, satisfying
\begin{equation}
\dot N = H .
\end{equation}
This means that $N$ is not introduced as a posterior bookkeeping device, but is solved dynamically together with $h$, $\dot h$, and $\rho_R$ from the background equations themselves. Numerically, the initial condition is chosen near the large field slow roll region, with a small initial radiation density and the corresponding warm slow roll velocity, and the coupled system is solved with an ODE integrator suitable for stiff equations.

At each integration step, the Hubble rate is obtained from the full Friedmann equation, while the temperature is inferred from the radiation energy density. 

The code therefore produces the full background trajectories $h(t)$, $\dot h(t)$, $\rho_R(t)$, and $N(t)$ directly, rather than imposing $h_*$, $Q_*$, or $N_{\rm end}$ in advance.

The end of inflation is determined by the event at which the full Hubble slow roll parameter first reaches unity. From the Friedmann equation and the background dynamics one obtains
\begin{equation}
\dot H=-\frac{\dot h^2+\frac43\rho_R}{2M_{\rm Pl}^2},\qquad
\epsilon_H\equiv-\frac{\dot H}{H^2}=
\frac{\dot h^2+\frac43\rho_R}{2M_{\rm Pl}^2H^2}.
\end{equation}
Unlike the potential slow roll approximation often used in cold inflation, this expression explicitly includes the contribution from the radiation energy density and therefore identifies the full dynamical end point of warm inflation. During the numerical integration, the condition
\begin{equation}
\epsilon_H=1
\end{equation}
defines the end time $t_{\rm end}$, from which one reads
\begin{equation}
N_{\rm end}=N(t_{\rm end}).
\end{equation}
Thus, $N_{\rm end}$ is not an external input but a derived quantity jointly by the full background evolution and the end of inflation event.

Once $N_{\rm end}$ has been determined, the pivot e-fold is defined by
\begin{equation}
N_*=N_{\rm end}-50 .
\end{equation}
The corresponding pivot time $t_*$ is obtained from the condition $N(t_*)=N_*$. At this point one reads off $h_*$, $\dot h_*$, $\rho_{R,*}$, $H_*$, $T_*$, and $\Upsilon_*$, and constructs
\begin{equation}
Q_* = \frac{\Upsilon_*}{3H_*},
\end{equation}
together with the channel fractions $f_{i,*}$ of the three dissipative components at the pivot scale. The $n_s$ and $r$ are then built from the potential slow roll quantities and the dissipative strength at the pivot scale and passed to the mock CMB or ACT lite profile scoring. The order of the present numerical pipeline is therefore as follows: first solve the full warm background trajectory, then determine $t_{\rm end}$, $N_{\rm end}$, and the pivot scale, and finally construct the warmness diagnostics and effective observables that enter the ranking.

\subsection{Numerical implementation, data term, and scoring prescription}

Two scoring layers are used. The first is a CAMB-based mock CMB likelihood, which takes $n_s$, $r$, and the scalar amplitude obtained from the warm proxy as input and performs a simplified comparison against the dominant information carried by the temperature power spectrum and lensing. The effective number of data points entering the BIC is written as $N=n_{\rm bin}\times n_{\rm spec}$. Under the default setting $n_{\rm bin}=98$ and $n_{\rm spec}=3$, the mock score therefore corresponds to $N=294$. The second layer is an optional ACT DR6 ACT-lite backend: when the relevant environment is available, the best candidates of the dissipative channels are rescored through a profile $\chi^2$. The expression ``ACT-lite backend quantity'' below refers to the additional score obtained by profiling over the ACT-lite sampling parameters at $(n_s,r)$.

After the unified scan has identified the best candidate point for each EDC, these points are subjected to a refined rescoring procedure in order to test how stable the hierarchy is against the diagnostic density. The term ``refined rescoring of points'' means that the parameters found in the unified scan are held while the common diagnostic sampling density is increased and the background trajectory, conservation error, warmness diagnostics, and backend score are recomputed. These rescored values are the ones quoted in the main tables and appendices.

For the statistical comparison, the Bayesian information criterion is adopted,
\begin{equation}
\mathrm{BIC}=\chi^2_{\min}+k\ln N,
\qquad
\Delta\mathrm{BIC}_i=\mathrm{BIC}_i-\mathrm{BIC}_{\mathrm{best}},
\end{equation}
where $k$ is the number of parameters effectively optimized for a given dissipative channel and $N$ is the effective number of data points used in the score. The role of the BIC here is to provide a transparent complexity penalty within a common scoring implementation, in order to assess whether a given channel gains enough statistical improvement to compensate for its extra freedom. It is not meant to replace a full Bayesian evidence calculation or a final model selection criterion.

\subsection{Self-consistency conditions, conditions not covered, and the scope of interpretation}

The conditions checked explicitly in the present work include the warmness requirement $T_*/H_*>1$ near the pivot scale, the dissipative strength $Q_*$, channel fractions, conservation error, and numerical computability. Stronger thermalization requirements, near-equilibrium conditions, and the impact of finite temperature corrections on the flatness of the potential have not been imposed point by point for every sample. The hierarchy derived below should therefore be understood first and foremost as a comparison among EDC within the present numerical setup, rather than as a final ranking obtained after all self-consistency conditions of warm inflation have been imposed pointwise.

Likewise, the matching to the standard post-inflationary thermal history is not imposed sample by sample as an independent criterion. The present analysis only requires that no explicit contradiction arises at the level of the model assumptions and the interpretation of the results. By compatibility with the standard thermal history, the paper means simply that no direct inconsistency has been identified under the current setup and checking level. The hierarchy given here therefore does not yet include any additional filtering induced by a full construction of the post-inflationary thermal history.

The results are thus best interpreted as a comparison of EDC based on warm background outputs and proxy effective observables, and should not be directly promoted to a final physical ranking of microscopic dissipative mechanisms.

\section{Results and discussion}\label{Sec4}

\subsection{Overall distribution of best fit points in $(n_s,r)$ plane}

\begin{figure}[H]
\centering
\includegraphics[width=0.7\textwidth, height=0.45\textheight, keepaspectratio]{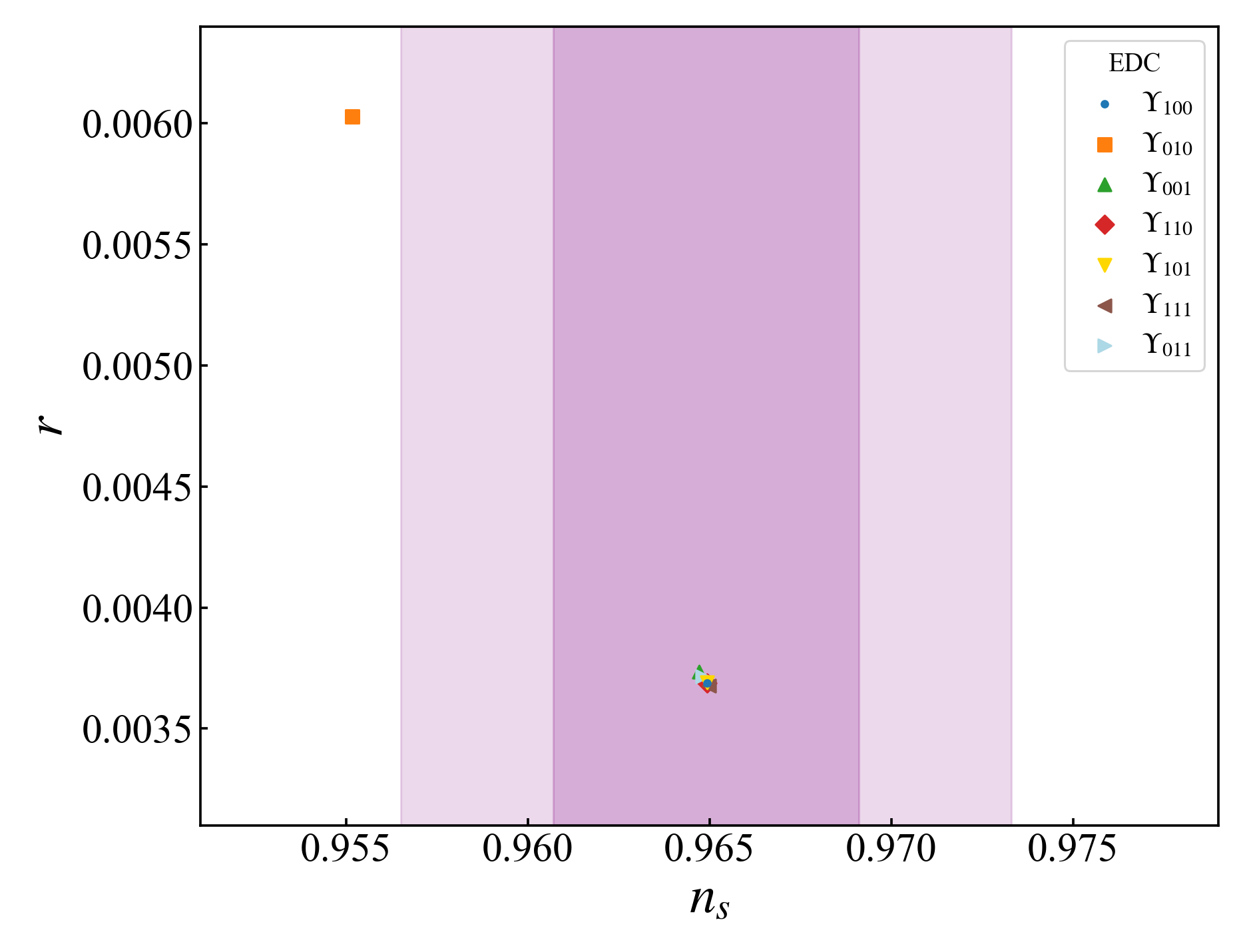}
\caption{Best fit points of the seven EDC in $(n_s,r)$ plane.}\label{nsr}
\end{figure}

As shown in Fig.~\ref{nsr}, the best fit points of the seven EDC exhibit a clear clustering pattern in $(n_s,r)$ plane. Except for the pure HT EDC $\Upsilon_{\mathrm{010}}$, the best fit points of the other six EDC are concentrated in a rather narrow region, with $n_s\approx0.965$ and $r\approx(0.003 \to 0.004)$. The main feature of the figure is that most EDC are not naturally separated in the most common two-dimensional observational projection, but instead appear strongly grouped together. By contrast, the best fit point of $\Upsilon_{\mathrm{010}}$ is significantly displaced from the main cluster, with smaller $n_s$ and larger $r$, thereby forming the clearest outlier in geometric terms.

Fig.~\ref{nsr} captures the overall geometry of the best fit point distribution, but not the final hierarchy itself. The more relevant physical issue is not which points happen to lie closer together in the plot, but how much complexity cost each EDC must pay to reach this observational region, which dominant channel it relies on, and which warmness regime it occupies. Fig.~\ref{nsr} alone is therefore insufficient for a ranking, and the $\Delta\mathrm{BIC}$ values, channel fractions, and warmness diagnostics must also be taken into account.

\subsection{$\Delta\mathrm{BIC}$ hierarchy}

\begin{figure}[H]
\centering
\includegraphics[width=0.70\textwidth, height=0.75\textheight, keepaspectratio]{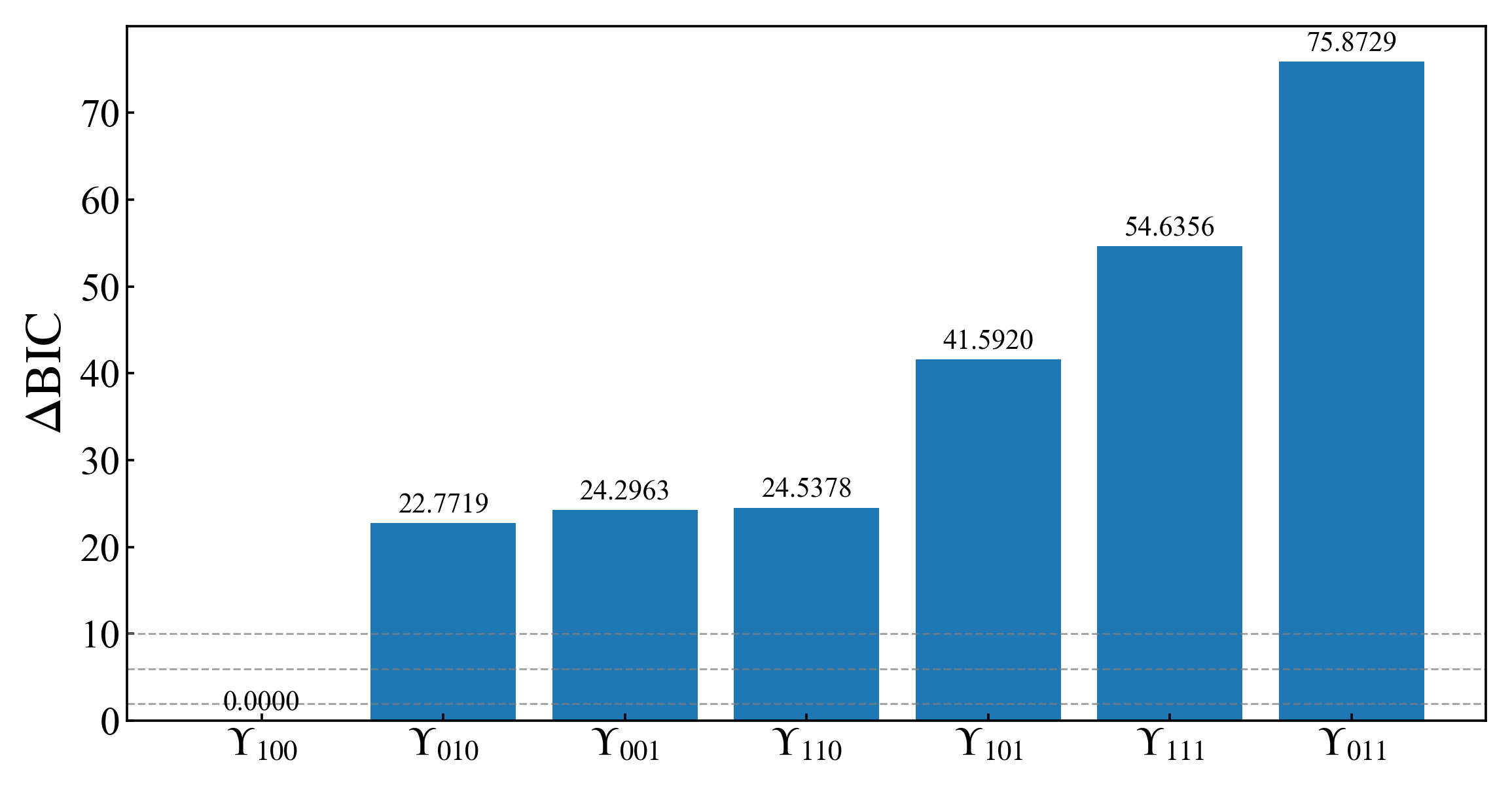}
\caption{$\Delta\mathrm{BIC}$ values relative to the top-ranked EDC.}\label{BIC}
\end{figure}

The statistical hierarchy among the EDC is more clearly seen from Fig.~\ref{BIC} and Tab.~\ref{Range}. The unified-search result refers to the first full scan over the seven channels under the common search boundaries listed in Tab.~\ref{Range}, the common scoring prescription, and the common diagnostic setup. Fig.~\ref{BIC} shows the $\Delta\mathrm{BIC}$ values of all channels relative to the best one, while Tab.~\ref{Range} lists both the unified-search score and the 1200-point refined rescoring.

\begin{table}[H]
\centering
\small
\caption{Relative BIC comparison between the unified scan and the 1200-point refined rescoring.}
\begin{tabular}{lcc}
\toprule
EDC & $\Delta\mathrm{BIC}$ (unified scan) & $\Delta\mathrm{BIC}$ (1200-point rescoring)\\
\midrule
$\Upsilon_{\mathrm{100}}$ & 0.0000 & 0.0000 \\
$\Upsilon_{\mathrm{010}}$ & 22.7719 & 29.9407 \\
$\Upsilon_{\mathrm{001}}$ & 24.2963 & 35.6882 \\
$\Upsilon_{\mathrm{110}}$ & 24.5378 & 33.0257 \\
$\Upsilon_{\mathrm{101}}$ & 41.5920 & 56.5016 \\
$\Upsilon_{\mathrm{111}}$ & 54.6356 & 76.7909 \\
$\Upsilon_{\mathrm{011}}$ & 75.8729 & 70.3350 \\
\bottomrule\label{Tab2}
\end{tabular}
\end{table}

This hierarchy has a direct physical interpretation. The advantage of the pure LT EDC $\Upsilon_{\mathrm{100}}$ is not a marginal fluctuation, but the clearest statistical preference that survives the complexity penalty. At the same time, multi-channel EDC do not automatically benefit from being more general: if the additional channels do not produce enough improvement to match their extra freedom, the complexity penalty pushes them back into disfavour.

From this point of view, the advantage of $\Upsilon_{\mathrm{100}}$ does not come from a special location in $(n_s,r)$ plane, but from the fact that it reaches an acceptable observational region with fewer degrees of freedom while maintaining a stable warmness range. In the present scoring setup, the pure LT channel already carries the main burden of the fit, so that $\Upsilon_{\mathrm{100}}$ remains preferred both in the unified scan and in the refined rescoring. The main best fit parameters for all channels are listed in Appendix~\ref{Appendix_A} (Tab.~\ref{Tab5}), while the principal observables, warmness diagnostics, and channel fractions are summarized in Tab.~\ref{Tab3}.

\subsection{1200-point refined rescoring}

The purpose of the 1200-point refined rescoring is to test whether the hierarchy obtained from the unified scan depends on a coarse numerical diagnostic. The rescoring does not re-optimize the parameters. Instead, it takes the best fit points found in the unified scan and recomputes the background trajectory, backend score, consistency error, warmness diagnostics, and channel fractions under a denser common diagnostic setup. If the advantage of a given channel were merely a coarse-grid artifact, its relative position should become unstable in this refined rescoring. Conversely, if the top EDC and the overall ordering survive, the conclusion can be regarded as reasonably robust.

Tab.~\ref{Range} and Tab.~\ref{Tab2} show the changes induced by the 1200-point rescoring. The top-ranked effective dissipation channel remains $\Upsilon_{\mathrm{100}}$, and the weakest EDC still gather at the tail, namely $\Upsilon_{\mathrm{011}}$ and $\Upsilon_{\mathrm{111}}$. The main shift appears in the middle layer: $\Upsilon_{\mathrm{110}}$ becomes slightly better than $\Upsilon_{\mathrm{001}}$, so that their relative order is exchanged, while $\Upsilon_{\mathrm{010}}$ still stays in the second position. In other words, the refined rescoring does not overturn the original ordering, but only separates a few EDC that were already close to one another in the unified scan. Numerically, the rescored $\Delta\mathrm{BIC}$ values become about $29.9$, $33.0$, $35.7$, and $56.5$ for $\Upsilon_{\mathrm{010}}$, $\Upsilon_{\mathrm{110}}$, $\Upsilon_{\mathrm{001}}$, and $\Upsilon_{\mathrm{101}}$, while $\Upsilon_{\mathrm{011}}$ and $\Upsilon_{\mathrm{111}}$ remain as large as $70.3$ and $76.8$.

Tab.~\ref{Tab2} therefore does not identify a new optimal effective dissipation channel. Rather, it shows that the hierarchy found in the unified scan remains stable under a finer diagnostic setting. Physically, the advantage of $\Upsilon_{\mathrm{100}}$ is not the result of a one-off fluctuation in the unified scan score. The changes occur mainly among a few EDC that were already close in the intermediate layer. The refined rescoring does not rewrite the main conclusion of the paper; it provides a sharper robustness check on the unified scan hierarchy.

\begin{table}[H]
\centering
\caption{Main results after the 1200-point refined rescoring.}
\begin{adjustbox}{max width=\textwidth}
\begin{tabular}{lccccccc}
\toprule
EDC & $\Delta\mathrm{BIC}$ & $n_s$ & $r$ & $Q_*$ & $T_*/H_*$ & max(cons err) & $N_*$ \\
\midrule
$\Upsilon_{\mathrm{100}}$ & $0.0000$ & $0.9649$ & $0.0037$ & $35.6851$ & $1903.2600$ & $3.8711\times10^{-4}$ & $99.2132$ \\
$\Upsilon_{\mathrm{010}}$ & $29.9407$ & $0.9552$ & $0.0060$ & $1.0023\times10^{-5}$ & $11.2034$ & $1.1156\times10^{-3}$ & $10.9088$ \\
$\Upsilon_{\mathrm{001}}$ & $35.6882$ & $0.9647$ & $0.0037$ & $1.1050$ & $6.0978$ & $2.1724\times10^{-3}$ & $59.5260$ \\
$\Upsilon_{\mathrm{110}}$ & $33.0257$ & $0.9649$ & $0.0037$ & $35.7644$ & $1872.1700$ & $3.9421\times10^{-4}$ & $99.7799$ \\
$\Upsilon_{\mathrm{101}}$ & $56.5016$ & $0.9649$ & $0.0037$ & $35.6847$ & $1900.4800$ & $3.8810\times10^{-4}$ & $99.2581$ \\
$\Upsilon_{\mathrm{111}}$ & $76.7909$ & $0.9650$ & $0.0037$ & $34.8460$ & $555.3840$ & $7.7680\times10^{-4}$ & $70.4464$ \\
$\Upsilon_{\mathrm{011}}$ & $70.3350$ & $0.9648$ & $0.0037$ & $1.6850$ & $0.3125$ & $1.4515\times10^{-3}$ & $89.8346$ \\
\bottomrule\label{Tab3}
\end{tabular}
\end{adjustbox}
\end{table}

\subsection{Channel fractions at the pivot scale}

\begin{figure}[H]
\centering
\includegraphics[width=0.70\textwidth, height=0.75\textheight, keepaspectratio]{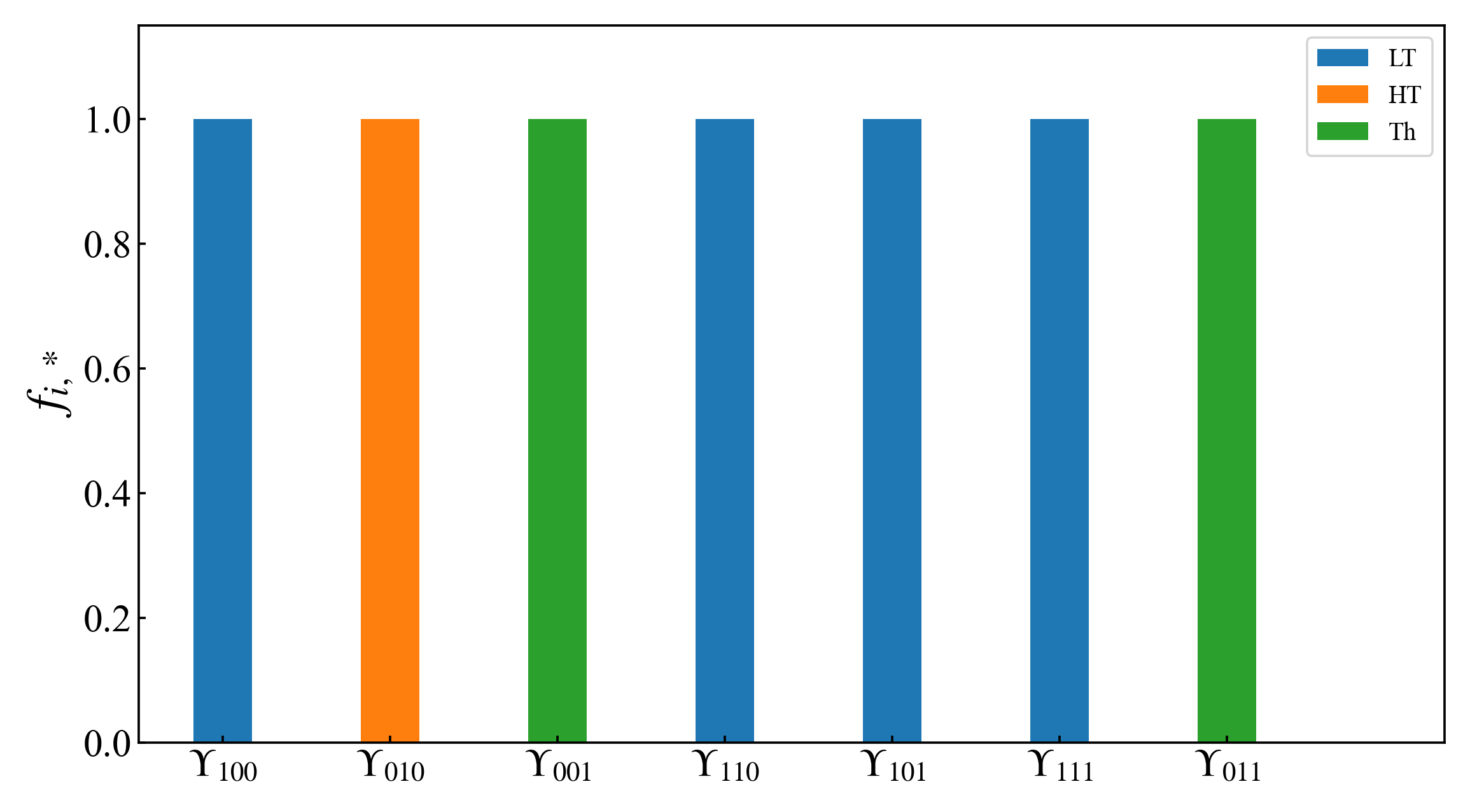}
\caption{Fractions of the three channels at the pivot scale, $f_{i,*}=\Upsilon_i/\Upsilon_{\mathrm{tot}}$.}\label{Fig3}
\end{figure}

Fig.~\ref{Fig3} shows the fractions $f_{i,*}=\Upsilon_i/\Upsilon_{\mathrm{tot}}$ of the three channels at the pivot scale. These fractions directly reveal the actual operating mode of each best fit point. The pattern is very clear. The three pure channel EDC correspond to clean single channel dominance: $\Upsilon_{\mathrm{100}}$ is LT dominated, $\Upsilon_{\mathrm{010}}$ is HT dominated, and $\Upsilon_{\mathrm{001}}$ is threshold-dominated. More importantly, the EDC that are nominally multi-channel do not generically exhibit a genuinely mixed dominant pattern near their best fit points: $\Upsilon_{\mathrm{110}}$, $\Upsilon_{\mathrm{101}}$, and $\Upsilon_{\mathrm{111}}$ all fall back to regions where the LT fraction is essentially unity, whereas the best fit point of $\Upsilon_{\mathrm{011}}$ lies almost completely in the threshold dominated region.

The physical significance of Fig.~\ref{Fig3} is that a multi-channel label does not automatically imply effective mixing. For the present set of EDC, one dominant channel is often enough to produce an acceptable spectral index and warmness range, while the remaining channels act only as weak corrections and do not form a stable, indispensable cooperative contribution.

\subsection{Warmness diagnostics: dissipative strength and thermal bath range}

\begin{figure}[H]
\centering
\includegraphics[width=0.70\textwidth, height=0.75\textheight, keepaspectratio]{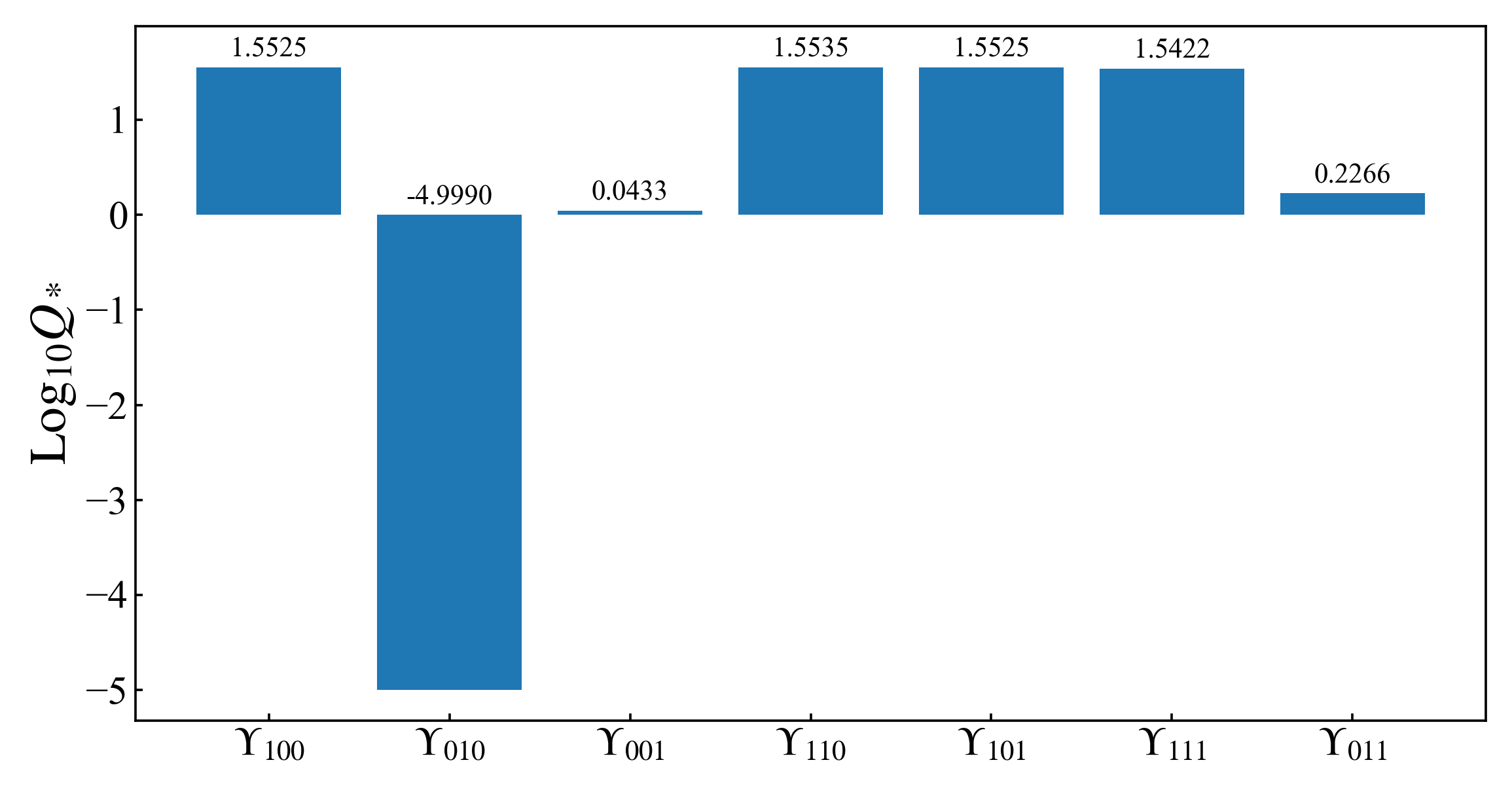}
\caption{$\mathrm{Log}_{10}Q_*$ at the best fit points of the seven EDC.}\label{Fig4}
\end{figure}

\begin{figure}[H]
\centering
\includegraphics[width=0.68\textwidth, height=0.7\textheight, keepaspectratio]{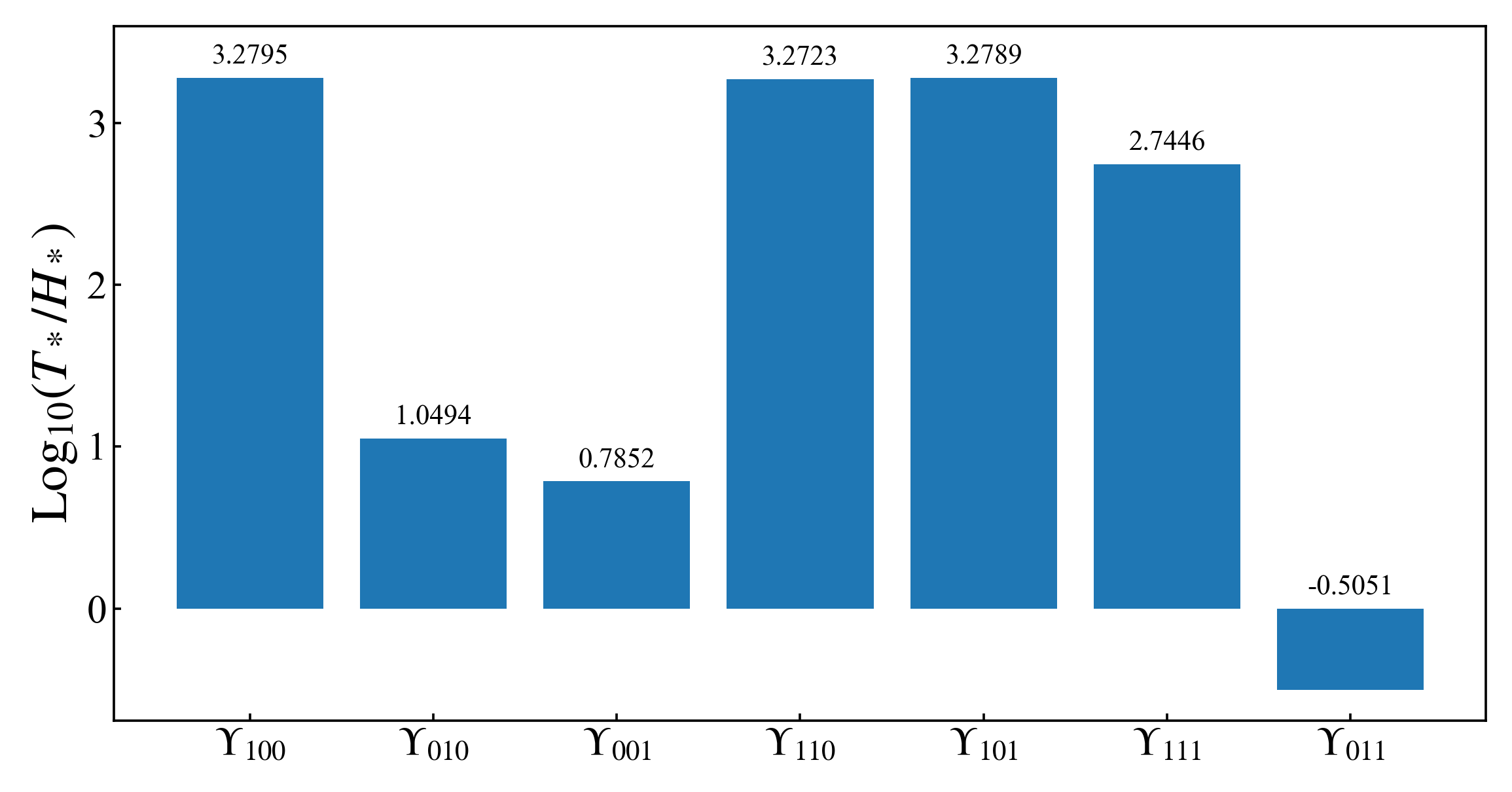}
\caption{$\mathrm{Log}_{10}(T_*/H_*)$ at the best fit points of the seven EDC.}\label{Fig5}
\end{figure}

Fig.~\ref{Fig4} and Fig.~\ref{Fig5}, together with Tab.~\ref{Tab3}, summarize the warmness diagnostics of the seven channels. Fig.~\ref{Fig4} shows that $\Upsilon_{\mathrm{100}}$, $\Upsilon_{\mathrm{110}}$, $\Upsilon_{\mathrm{101}}$, and $\Upsilon_{\mathrm{111}}$ occupy the high-$\log_{10}Q_*$ region and therefore correspond to stronger dissipation; $\Upsilon_{\mathrm{001}}$ and $\Upsilon_{\mathrm{011}}$ are noticeably lower; $\Upsilon_{\mathrm{010}}$ lies at the weakest dissipation end. Fig.~\ref{Fig5} gives the relative positions in $T_*/H_*$. Again, $\Upsilon_{\mathrm{100}}$, $\Upsilon_{\mathrm{110}}$, and $\Upsilon_{\mathrm{101}}$ lie in the higher range, while $\Upsilon_{\mathrm{111}}$ still preserves warmness but at a lower level than the first three. The statistical best fit point of $\Upsilon_{\mathrm{011}}$ has already crossed below the warmness threshold.

\begin{table}[H]
\centering
\caption{Observables, warmness diagnostics, and channel fractions at the best fit points of the seven EDC.}
\begin{adjustbox}{max width=\textwidth}
\begin{tabular}{lcccccccc}
\toprule
EDC & $n_s$ & $r$ & $Q_*$ & $T_*/H_*$ & $\Delta\mathrm{BIC}$ (unified scan) & $f_{\rm LT,*}$ & $f_{\rm HT,*}$ & $f_{\rm Th,*}$ \\
\midrule
$\Upsilon_{\mathrm{100}}$ & $0.9649$ & $0.0037$ & $35.6851$ & $1903.2600$ & $0.0000$ & $1.0000$ & $0.0000$ & $0.0000$ \\
$\Upsilon_{\mathrm{010}}$ & $0.9552$ & $0.0060$ & $1.0023\times10^{-5}$ & $11.2034$ & $22.7719$ & $0.0000$ & $1.0000$ & $0.0000$ \\
$\Upsilon_{\mathrm{001}}$ & $0.9647$ & $0.0037$ & $1.1050$ & $6.0978$ & $24.2963$ & $0.0000$ & $0.0000$ & $1.0000$ \\
$\Upsilon_{\mathrm{110}}$ & $0.9649$ & $0.0037$ & $35.7644$ & $1872.1700$ & $24.5378$ & $1.0000$ & $0.0000$ & $0.0000$ \\
$\Upsilon_{\mathrm{101}}$ & $0.9649$ & $0.0037$ & $35.6847$ & $1900.4800$ & $41.5920$ & $1.0000$ & $0.0000$ & $0.0000$ \\
$\Upsilon_{\mathrm{111}}$ & $0.9650$ & $0.0037$ & $34.8460$ & $555.3840$ & $54.6356$ & $1.0000$ & $1.0041\times10^{-7}$ & $0.0000$ \\
$\Upsilon_{\mathrm{011}}$ & $0.9648$ & $0.0037$ & $1.6850$ & $0.3125$ & $75.8729$ & $0.0000$ & $7.9185\times10^{-8}$ & $1.0000$ \\
\bottomrule\label{Tab4}
\end{tabular}
\end{adjustbox}
\end{table}

These results show that the differences among the channels lie mainly in the way the best fit points are realized, rather than in the locations of the points themselves. The EDC $\Upsilon_{\mathrm{100}}$, $\Upsilon_{\mathrm{110}}$, and $\Upsilon_{\mathrm{101}}$ reach similar $(n_s,r)$ values while remaining in a regime of comparatively strong dissipation and high thermal-bath strength. By contrast, $\Upsilon_{\mathrm{010}}$ and $\Upsilon_{\mathrm{011}}$ are closer to marginal operating regions: the former corresponds to extremely weak dissipation, while the latter has already crossed below the warmness threshold.

The case $\Upsilon_{\mathrm{010}}$ indicates that a pure HT channel tends to be pushed into a more marginal regime if it is to maintain a locally acceptable fit, whereas the statistical best fit point of $\Upsilon_{\mathrm{011}}$ already lies below the warmness threshold. It is therefore more appropriate to treat $\Upsilon_{\mathrm{011}}$ as a reference point near the edge of the present parameterization than to interpret it on exactly the same footing as the EDC that remain securely in the warm regime.

\subsection{Boundary checks and internal mixing probes}\label{4.6}

Fig.~\ref{Fig4} and Fig.~\ref{Fig5}, Tab.~\ref{Tab3}, Tab.~\ref{Tab4} and Appendix~\ref{Appendix_B} show that the best fit points of the multi-channel EDC still carry two or three active channels at the level of labels, but their actual contributions at the pivot scale are often already concentrated in a single dominant channel. To determine whether this is merely a local feature near the best fit point, boundary checks and constrained internal-mixing probes were performed.

The paper defines an ``internally mixed region'' by requiring that each nominally active channel satisfy
\begin{equation}
f_{i,*}\ge 0.05 .
\end{equation}
This threshold separates genuine participation of a channel from cases where the channel is retained in the label but its actual contribution is already negligible. Tab.~\ref{Tab3} shows that the HT fractions of $\Upsilon_{\mathrm{111}}$ and $\Upsilon_{\mathrm{011}}$ are far below $0.05$, while the best fit points of $\Upsilon_{\mathrm{110}}$ and $\Upsilon_{\mathrm{101}}$ have already returned to the single-dominant boundary $f_{\rm LT,*}=1$. In other words, the activation of multiple channels in the present optimum acts more as a formal label than as a stable internally cooperative working regime; the fact that $\Upsilon_{\mathrm{011}}$ also has $T_*/H_*<1$ reinforces that it lies near a weakly warm edge.

The boundary checks show that this pull toward the boundary is reasonably stable. In Appendix~\ref{Appendix_B}, $\Upsilon_{\mathrm{110}}$, $\Upsilon_{\mathrm{101}}$, and $\Upsilon_{\mathrm{111}}$ are all identified as LT dominated in both the unified scan and the 1200-point refined rescoring, while $\Upsilon_{\mathrm{011}}$ remains threshold-dominated. The identity of the dominant channel is therefore not materially altered by the change in diagnostic density.

This conclusion is further supported by Tab.~\ref{Tab6} in Appendix~\ref{Appendix_B}, while the classification of the boundary positions near the best fit points is summarized in Tab.~\ref{Tab7} of Appendix~\ref{Appendix_C}. For $\Upsilon_{\mathrm{101}}$, imposing the condition $f_{i,*}\ge0.05$ raises the penalized objective function from $1061.535$ to $1209.353$, and the returned channel fractions still lie near the LT dominated boundary. The other numerical results are listed in Tab.~\ref{Tab6}. Taken together, Tab.~\ref{Tab6} and Tab.~\ref{Tab7} show that, once the solution is pushed away from the boundary toward an internally mixed region, the objective function generally becomes worse and some EDC no longer admit a stable and viable internally mixed solution. Under the present parameterization and scoring setup, the best fit points of the multi-channel EDC therefore do not form a robust internally mixed region, but more stably return to LT-dominated or threshold-dominated boundaries.

\section{Conclusions}\label{Sec5}

A unified comparison has been carried out for seven EDC built from the LT, HT, Th channels within the warm Higgs plateau potential. The comparison is performed under a common warm-background solver, a common complexity penalty, and a common boundary check criterion, with the aim of determining whether these EDC admit a stable and physically interpretable hierarchy when examined on the same footing.

The best fit points do not separate naturally in ($n_s$, $r$) plane. With the exception of the pure high-temperature effective dissipation channel $\Upsilon_{\mathrm{010}}$, the other six EDC remain concentrated in a narrow region around $n_s \approx 0.965$ and $r \approx (3.68$--$3.74)\times10^{-3}$. By contrast, $\Upsilon_{\mathrm{010}}$ is clearly displaced from this main cluster, with a representative best fit point near $n_s = 0.95517$ and $r = 6.03\times10^{-3}$. This indicates that the two-dimensional location in ($n_s$, $r$) plane is not, by itself, sufficient to determine the hierarchy among the dissipative channels.

Under both the unified scan and the 1200-point refined rescoring, the pure LT EDC $\Upsilon_{\mathrm{100}}$ remains top-ranked. After rescoring, the corresponding $\Delta \mathrm{BIC}$ values are about $29.94$ for $\Upsilon_{\mathrm{010}}$, $33.03$ for $\Upsilon_{\mathrm{110}}$, $35.69$ for $\Upsilon_{\mathrm{001}}$, $56.50$ for $\Upsilon_{\mathrm{101}}$, $70.34$ for $\Upsilon_{\mathrm{011}}$, and $76.79$ for $\Upsilon_{\mathrm{111}}$, showing that the overall hierarchy is stable under the present boundary check criterion. This statistical ordering is consistent with the physical working regime of each EDC. The best fit point of $\Upsilon_{\mathrm{100}}$ gives $Q_* \approx 35.685$ and $T_*/H_* \approx 1.90\times10^{3}$, placing it in a clearly strong warm regime, whereas $\Upsilon_{\mathrm{011}}$ yields $T_*/H_* \approx 0.312$, already below the warmness threshold. In this sense, the advantage of $\Upsilon_{\mathrm{100}}$ lies not only in its ranking, but also in a clearer and more robust physical operating regime.

The channel fractions, boundary checks, and constrained internal-mixing probes show that the best fit points of the multi-channel EDC do not form a stable internally mixed region. Instead, they are generally pulled back toward a single dominant channel. In particular, $\Upsilon_{\mathrm{110}}$, $\Upsilon_{\mathrm{101}}$, and $\Upsilon_{\mathrm{111}}$ are all LT-dominated, while $\Upsilon_{\mathrm{011}}$ remains close to the threshold-dominated corner. This shows that, under the present setup, the hierarchy among the dissipative channels is controlled less by their positions in ($n_s, r$) plane than by the complexity cost required to reach those positions, the dominance pattern near the best fit point, and the warmness range that can be maintained.

The present results show that combining dissipative channel comparisons with boundary checks shifts the comparison of different dissipative forms from best fit point locations alone to the level of dominance pattern and stability. Since the conclusions are still derived within a framework based on warm-background outputs and proxy effective observables, they are more appropriately understood as a comparison of EDC under the present unified implementation, rather than as a final judgement on the underlying microscopic dissipative mechanisms.

\hspace{2cm}
\section*{Acknowledgements}
Wei Cheng was supported by Chongqing Natural Science Foundation project under Grant No. CSTB2022NSCQ-MSX0432, by Science and Technology Research Project of Chongqing Education Commission under Grant No. KJQN202200621, and by Chongqing Human Resources and Social Security Administration Program under Grants No. D63012022005. 
Jia-Wei Zhang is supported by National Natural Science Foundation of China under Contract No.12275036, the Natural Science Foundation of Chongqing under Contract  No. CSTB2025NSCQ-GPX0945, cstc2021jcyjmsxmX0681.
Ruiyu Zhou is supported by the National Natural Science Foundation of China under Contract No. 12305109, and No. 12147102, Chongqing Natural Science Foundation project under Grant No. CSTB2022NSCQ-MSX0534, and Science and Technology Research Project of Chongqing Municipal Education Commission under Grant No. KJQN202300614.

\section*{Appendix}

\appendix
\refstepcounter{section}
\subsection*{Appendix \thesection: Main parameter values at the best fit points}
\label{Appendix_A}

Tab.~\ref{Tab5} lists those main parameters that participate in the calculation at the best fit point of each EDC. Parameters associated with channels that are not switched on are denoted uniformly by ``--''.

\begin{table}[H]
\centering
\caption{Main parameter values at the best fit points of the seven EDC.}
\begin{adjustbox}{max width=\linewidth}
\begin{tabular}{lccccccc}
\toprule
Parameter &$\Upsilon_{\mathrm{100}}$ & $\Upsilon_{\mathrm{010}}$ & $\Upsilon_{\mathrm{001}}$ & $\Upsilon_{\mathrm{110}}$ & $\Upsilon_{\mathrm{101}}$ & $\Upsilon_{\mathrm{111}}$ & $\Upsilon_{\mathrm{011}}$ \\
\midrule
$\lambda$ & 0.0798 & 0.0165 & $1.0000 \times 10^{-7}$ & 0.0084 & 0.0253 & 0.0658 & 0.0224 \\
$\xi$ & 8180210.0000 & 364484.0000 & 1.0000 & 2555740.0000 & 4587900.0000 & 642777.0000 & 1.0000 \\
$C_{\rm LT}$ & $1.0000 \times 10^{8}$ & - & - & $1.0000 \times 10^{8}$ & $1.0000 \times 10^{8}$ & $3.5617 \times 10^{6}$ & - \\
$C_{\rm HT}$ & - & $3.1319 \times 10^{-9}$ & - & $1.0000 \times 10^{-12}$ & - & $4.3755 \times 10^{-9}$ & $9.9100 \times 10^{-9}$ \\
$C_{\rm Th}$ & - & - & $3.9935 \times 10^{7}$ & - & 257.6640 & $1.0000 \times 10^{8}$ & 0.3960 \\
$g$ & - & - & $1.0000 \times 10^{-5}$ & - & 1.3497 & 0.3471 & $1.0000 \times 10^{-5}$ \\
$a$ & - & - & 6.0000 & - & 4.0579 & 6.0000 & 4.5557 \\
$c$ & - & - & 2.3943 & - & 1.2323 & 2.2985 & 2.2696 \\
$\alpha_h$ & - & 0.2274 & - & -1.1355 & - & 1.2000 & 1.2000 \\
$\alpha_T$ & - & -1.2000 & - & 1.0655 & - & -0.7389 & -0.5442 \\
$h_0$ & 0.4059 & 0.0025 & - & 0.4104 & 0.4064 & 0.0313 & 0.0004 \\
$\mu$ & - & 0.0001 & 5.6202 & 0.0001 & 0.0002 & 1.7543 & 0.2858 \\
$m_0$ & - & - & 0.0010 & - & 1.3733 & 0.0016 & 0.0187 \\
$\kappa_T$ & - & - & 10.1346 & - & 17.2879 & 23.7425 & 17.6683 \\
\bottomrule\label{Tab5}
\end{tabular}
\end{adjustbox}
\end{table}

\refstepcounter{section}
\subsection*{Appendix \thesection: Boundary checks and constrained probes}
\label{Appendix_B}

Tab.~\ref{tab:boundary_summary} summarizes the main numerical outputs of the boundary checks and the constrained internal mixing probes; the interpretation is given in Sec.~\ref{4.6}.

\begin{table}[H]
\centering
\caption{Summary of boundary checks and constrained probes.}
\footnotesize
\setlength{\tabcolsep}{4pt}
\begin{tabular}{
>{\raggedright\arraybackslash}m{0.27\textwidth}
>{\centering\arraybackslash}m{0.16\textwidth}
>{\centering\arraybackslash}m{0.16\textwidth}
>{\centering\arraybackslash}m{0.16\textwidth}
>{\centering\arraybackslash}m{0.16\textwidth}
}
\toprule
Item & $\Upsilon_{\mathrm{110}}$ & $\Upsilon_{\mathrm{101}}$ & $\Upsilon_{\mathrm{111}}$ & $\Upsilon_{\mathrm{011}}$ \\
\midrule

Stably close to boundary?
& Yes & Yes & Yes & Yes \\
\addlinespace[2pt]

Dominant channel
& \makecell{unified scan:\\LT\\1200-point:\\rescoring: LT}
& \makecell{unified scan:\\LT\\1200-point:\\rescoring: LT}
& \makecell{unified scan:\\LT\\1200-point:\\rescoring: LT}
& \makecell{unified scan:\\Th\\1200-point:\\rescoring: Th} \\
\addlinespace[2pt]

Baseline penalized\\objective
& $1.4730 \times 10^{4}$
& 1061.5350
& $1.0000 \times 10^{9}$
& 239.1320 \\
\addlinespace[2pt]

Constrained\\penalized objective
& $1.0000 \times 10^{9}$
& 1209.3530
& $1.0000 \times 10^{9}$
& $1.0000 \times 10^{9}$ \\
\addlinespace[2pt]

Difference
& $1.0000 \times 10^{9}$
& $1.4782 \times 10^{2}$
& 0.0000
& $1.0000 \times 10^{9}$ \\
\addlinespace[2pt]

Returned $f_*$ in\\constrained probe
& --
& $(1,0,6.1780 \times 10^{-308})$
& --
& -- \\
\addlinespace[2pt]

Interpretation
& \small No stable viable mixed solution; strong boundary attraction
& \small Finite constrained value obtained; back to LT boundary
& \small Baseline hit upper bound
& \small No viable two-channel solution \\

\bottomrule\label{Tab6}
\end{tabular}
\label{tab:boundary_summary}
\end{table}

\noindent\textit{Note}: the ``baseline penalized objective'' is the penalized objective function for the boundary type baseline point, and the ``constrained penalized objective'' is the result after imposing the minimum channel-fraction constraint. For $\Upsilon_{\mathrm{110}}$ and $\Upsilon_{\mathrm{011}}$, the constrained value directly reaches the preset upper bound, indicating that once the solution is pushed away from the boundary toward the internally mixed region, the current probe no longer maintains a stable viable solution. For $\Upsilon_{\mathrm{101}}$, the constrained value is finite, but the result still contracts back to the LT dominated boundary. For $\Upsilon_{\mathrm{111}}$, the present probe setup does not support a stronger conclusion.

\refstepcounter{section}
\subsection*{Appendix \thesection: Boundary properties near the best fit points}
\label{Appendix_C}

Tab.~\ref{tab:boundary_properties} lists the classification results for the boundary properties near the best fit points; the related discussion is given in Sec.~\ref{4.6}.

\newcolumntype{C}[1]{>{\centering\arraybackslash}m{#1}}
\newcolumntype{L}[1]{>{\raggedright\arraybackslash}m{#1}}
\begin{table}[H]
\centering
\caption{Boundary properties near the best fit points and their interpretation.}
\scriptsize
\setlength{\tabcolsep}{3.5pt}
\begin{tabular}{
  L{1.9cm}
  L{2.2cm}
  L{2.3cm}
  C{1.7cm}
  C{2.0cm}
  L{4.7cm}
}
\toprule
Channel\slash parameter & Current status & Boundary position & Boundary type & Relation to main conclusion & Interpretation \\
\midrule
$\Upsilon_{\mathrm{110}}$ best fit point
& $f_{\rm LT,*}\approx1$; close to boundary in unified scan \& 1200-point rescoring
& LT-dominated boundary submanifold
& structural
& direct
& Both boundary checks and constrained probe show LT-dominated boundary is strongly attractive; no stable internally mixed region under present setup. \\
\addlinespace[2pt]
$\Upsilon_{\mathrm{101}}$ best fit point
& $f_{\rm LT,*}\approx1$; constrained-probe $\Delta \approx 1.4782 \times 10^{2}$
& LT-dominated boundary submanifold
& structural
& direct
& After internal mixing, objective function deteriorates markedly; result describes boundary preference, not strict exclusion. \\
\addlinespace[2pt]
$\Upsilon_{\mathrm{111}}$ best fit point
& $f_{\rm LT,*}\approx1$; stably close to boundary under checks
& LT-dominated boundary submanifold
& structural
& cautiously direct
& best fit remains near boundary, but baseline/constrained values hit upper bound; no stronger physical reading taken. \\
\addlinespace[2pt]
$\Upsilon_{\mathrm{011}}$ best fit point
& $f_{\rm Th,*}\approx1$; close to boundary in unified scan \& rescoring
& threshold-dominated boundary submanifold
& structural
& direct
& HT--Th double-channel does not form stable mixed region near optimum; returns to threshold-dominated corner. \\
\addlinespace[2pt]
$\Upsilon_{\mathrm{100}}$: $C_{\rm LT}$
& $1.0000 \times 10^{8}$
& coincident with displayed upper boundary
& continuous parameter
& indirect
& Extreme LT-amplitude limit; does not alter $\Upsilon_{\mathrm{100}}$ as top-ranked effective dissipation channel. \\
\addlinespace[2pt]
$\Upsilon_{\mathrm{010}}$: $\alpha_T,\mu$
& $\alpha_T=-1.2000$; $\mu=1.0000 \times 10^{-4}$
& coincident with confirmed lower boundaries
& continuous parameter
& indirect
& Weak identification in HT temperature-power \& reference-scale; boundary-sensitive directions. \\
\addlinespace[2pt]
$\Upsilon_{\mathrm{001}}$: $\lambda,\xi,g,a,\alpha_T,m_0$
& $1.0000 \times 10^{-7}$; $1.0000$; $1.0000 \times 10^{-5}$; $6.0000$; $1.2000$; $1.0000 \times 10^{-3}$
& coincident with displayed/confirmed boundaries
& continuous parameter
& indirect
& Pure threshold solution relies on extreme parameters; does not alter $\Upsilon_{\mathrm{001}}$ effective dissipation channel conclusion. \\
\addlinespace[2pt]
$\Upsilon_{\mathrm{110}}$: $C_{\rm LT}, C_{\rm HT}, c, \mu$
& $1.0000 \times 10^{8}$; $1.0000 \times 10^{-12}$; $-2.0000$; $1.0000 \times 10^{-4}$
& coincident with displayed/confirmed boundaries
& continuous parameter
& indirect
& Extra channel suppressed/compensated; consistent with contraction to LT-dominated boundary. \\
\addlinespace[2pt]
$\Upsilon_{\mathrm{101}}$: $C_{\rm LT}, C_{\rm Th}, \alpha_h$
& $1.0000 \times 10^{8}$; $1.0000 \times 10^{8}$; $1.2000$
& coincident with displayed upper boundaries
& continuous parameter
& indirect
& Amplitude/power-law directions act as compensators; no effective mixed dominance pattern. \\
\addlinespace[2pt]
$\Upsilon_{\mathrm{111}}$: $a,\alpha_h$
& $6.0000$; $1.2000$
& coincident with confirmed upper boundaries
& continuous parameter
& cautiously indirect
& Boundary-sensitive directions; $\Upsilon_{\mathrm{111}}$ assessment governed by collapse to LT dominance. \\
\addlinespace[2pt]
$\Upsilon_{\mathrm{011}}$: $\xi,g,\alpha_h$
& $1.0000$; $1.0000 \times 10^{-5}$; $1.2000$
& coincident with displayed/confirmed boundaries
& continuous parameter
& indirect
$\Upsilon_{\mathrm{011}}$ best fit near threshold-dominated corner with boundary saturation of HT/threshold parameters. \\
\bottomrule\label{Tab7}
\end{tabular}
\label{tab:boundary_properties}
\end{table}

\end{document}